\documentclass[twocolumn,superscriptaddress,floatfix,showpacs,aps]{revtex4}
\usepackage{graphicx,amsfonts,amsmath,bm,color,tabularx,epstopdf}
\usepackage{multirow}
\usepackage[normalem]{ulem}

\begin{document}

\title{A new approach to the Kondo effect: sophomore's dream function and Lee-Yang zeros}

\author{Valentin Yu. Irkhin}
\affiliation{M. N. Mikheev Institute of Metal Physics, S. Kovalevskaya str. 18, 620077 Ekaterinburg, Russia}

\date{\today}

\begin{abstract}
We develop an analytical approach to the Kondo problem based on an effective partition function that generalizes the sophomore’s dream integral. The Kondo crossover is represented as a passage through an approximately equidistant  ladder of its Lee–Yang zeros in the complex  plane, rather than as a singularity on the real axis.
Summation over this ladder yields the Gamma function  $\Gamma(\alpha + T_K/(2 \pi T))$ as the universal crossover backbone, where the offset $\alpha =1/2$ is fixed by the Friedel sum rule. 
The Euler–Maclaurin expansion connects the discrete zero structure to the  Nozi\`eres Fermi-liquid theory through Bernoulli numbers. A two-cutoff  formula provides a compact, essentially one-parameter description of the full crossover in terms of a  temperature-space Fano lineshape which is a thermal average of the energy-space Fano resonance observed in STM.
The approach  extends to $\mathrm{SU}(N)$ Coqblin-Schrieffer, underscreened and Anderson impurity models, and Kondo lattices. 
\end{abstract}
\maketitle

\textit{Introduction}.
The Kondo effect \cite{Kondo,Hewson} -- the anomalous scattering of conduction electrons by magnetic impurities -- remains one of the paradigmatic problems of many-body physics, which has a long and dramatic history. Despite exact solutions via Bethe ansatz \cite{Andrei,Wiegmann} and powerful numerical renormalization group (NRG) methods \cite{Wilson,Bulla}), simple analytical formulas describing the full temperature crossover are still lacking. The Hamann formula \cite{Hamann} gives an interpolation description of resistivity $\rho(T)$ at intermediate temperatures, captures (with non-exact factors) the high-$T$ logarithmic behavior, but fails at low $T$. The Nozi\`eres Fermi liquid (FL) theory \cite{Nozieres} gives exact low-$T$ coefficients but does not extend to the crossover. 

We show that such an interpolation arises naturally from the partition function 
\begin{eqnarray}
Z(t,k,a) = \int_{0}^{\infty }\exp(te^{-ky}(y- a)-y)dy 
\end{eqnarray}
satisfying \cite{Gabor,Irkhin}
\begin{eqnarray}
Z(t,k,a) =\int_{0}^1 (e^ax)^{-tx^k} dx =\sum_{n=0}^{\infty} \frac{t^n  e_n(-(1 + nk) a)}{(1 + nk)^{n+1}}\label{som}
\end{eqnarray}
where $e_n(x) = \sum_{j=0}^n x^j/j!=e^x \Gamma(n+1,x))/n!$ is the exponential sum function, $e_n(0)=1$, $\Gamma(n,x)$ being the upper incomplete Gamma function.

At  $t= k =1, \, a = 0$, Eq. (\ref{som}) reduces to  the classical identity $$\int_0^1 x^{-x} dx = \sum^\infty_{n=0} n^{-n}=1.2913$$ discovered by Johann Bernoulli in 1697 \cite{Bernoulli} and later named ``sophomore's dream'' owing to its  counterintuitive simplicity.  The form $y \exp(-ky)$ is a simple universal function which
enables one to construct clear and solvable models for various physical problems in statistical physics and quantum field theory.

The generalized sophomore's dream partition function (1) has remarkably rich analytical properties (some results are presented briefly in Appendix): it is entire in $t$, has zeros with approximately periodic spacing $s = 2\pi ek$ in the complex plane and admits a Weierstrass (Hadamard) product representation \cite{Irkhin},
\begin{equation}
Z(t,k,a) = e^{h(t)}\prod_{n}^{}\left( 1 - \frac{t}{t_{n}} \right)e^{t/t_{n}}. \label{Had}
\end{equation}
with $h(t)$ an entire function. This factorization through zeros is the foundation for  applications below.
 The  complex-plane zeros can be identified by the Lee-Yang (Fisher) zeros \cite{Lee,Fisher,Boo}, which  provide description of phase transitions and crossovers  when the zeros move onto the real axis of the control parameter.

 {We stress that the use of Lee–Yang zeros here does not presuppose an underlying phase transition: analogous off-axis zero trajectories are used to characterize genuine crossovers (in the strict absence of any singularity) in other contexts, e.g., the QCD chiral crossover \cite{Stephanov,Ding}, and have been observed experimentally in finite (single-spin) quantum systems with no thermodynamic limit \cite{Boo}. For the physical ($a>0$) Kondo parameter range, the zeros of $Z(t.a)$ remain at a fixed finite distance from the real axis for all $t$; only for the unphysical $a<0$ do they approach the real axis, which is consistent with the fact that the Kondo temperature marks a crossover, not a transition.}

 In the present paper we apply the function (1) to describe various regimes in the Kondo model, especially the Kondo crossover.


\textit{Model and setup.}
We treat the Kondo model 
\begin{equation}
H = \sum_{\bf k\sigma}\varepsilon_{\bf k}c_{\bf k\sigma}^{\dagger}c_{\bf k\sigma} + J\,\mathbf{S} \cdot \mathbf{s}
\end{equation}
describing exchange interaction of conduction electrons with localized spin $S$.
This model demonstrates from weak-to-strong coupling crossover  characterized by the Kondo temperature $T_K \simeq D\exp(-1/g)$, with $g=2JN(E_F)$ the dimensionless coupling constant, $N(E_F)$ the density of electron states at the Fermi level. 
We put $k=1$ and write $Z(t,a) = Z(t,1,a)$. This choice is motivated by the requirement that the weak-coupling regime be governed by the standard Kondo scaling variable $g(T) \propto 1/\ln(T/T_K)$ and by integer powers of $g$ (possibly with $\ln\ln$ corrections), as in conventional perturbative RG,
and $a$ corresponds to  
symmetry violation by crystal field, magnetic field,  spin dynamics, etc.



The partition function $Z(t,a)$ should be viewed not as the exact  partition function of the Kondo problem, but as an effective collective-coordinate approximation to it. 
Rather than attempting to rederive the microscopic solution, we aim to identify the universal analytic structure underlying the Kondo crossover. The variable 
$y$ represents a reduced screening degree of freedom, while the physical RG flow enters through the temperature-dependent trajectory $t(T)$. In this sense, the present formalism is variational: the analytical structure of $Z$ is fixed by the Lee–Yang zero ladder and the required asymptotic properties, whereas the embedding of the true many-body problem is encoded in the choice of  $t(T)$. Different observables may then emphasize different asymptotic realizations of the same reduced partition function, while higher-order corrections may be incorporated as fluctuations around this collective description.

{The exponent of Eq. (1) can be treated as a kinetic ($-y$) plus potential ($te^{-y} (y-a)$) decomposition reminiscent of the Anderson–Yuval–Hamann Coulomb-gas representation of the Kondo problem \cite{Yuval}, with $y$ playing the role of the logarithmic RG ``time'' $\ln(D/E)$, $te^{-y}$ encoding the one-loop multiplicative renormalization of the coupling (effective exchange amplitude at the scale $E$).  $a\sim \ln(D/\Delta)$ represents the explicit symmetry-breaking field, so that the factor $y-a$  describes the detuning of this amplitude from a reference depth set by a competing energy scale $\Delta$, changing sign once the flow crosses $E\sim \Delta$.
Then $Z(t,a)$  corresponds to the free-fermion (Toulouse-point/non-interacting-charge) limit of the Coulomb gas; residual charge–charge (kink–kink) interactions, analogous to the thermodynamic Bethe ansatz (TBA)  kernel, are captured only approximately, as discussed below -- through the logarithmic drift corrections (see Eq. (\ref{drift}).}

We will  construct the scattering phase shift for the Kondo problem as a sum over the Lee-Yang zeros of $Z(t, a)$ both in the weak and strong coupling regime. In the latter case, this sum has the structure of a fermionic Matsubara frequency summation. 
The Euler-Maclaurin formula provides the exact bridge between the discrete Lee-Yang zeros and the continuous FL theory of Nozi\`eres. 
The Kondo crossover is the smooth passage of the RG trajectory $t(T)$ through the region of influence of the zero ladder, and not a Lee-Yang phase transition (in fact, in our model (1) the  Lee-Yang zeros can approach the real axis for \textit{negative} $a$ \cite{Irkhin} demonstrating the Stokes phenomenon). However, the language of Lee-Yang zeros is still useful because it organizes the perturbation theory and generates the formal Gamma function structure. 

\textit{Nearest-zero approach.}
The zeros of $Z(t,0)$ in the complex $t$-plane can be related to those of the error function and form approximately conjugate pairs \cite{Irkhin}:
\begin{equation}
t_n\simeq 1+e[-(1/2) \ln (2\pi^2(n-1/8)) +2\pi i(n-1/8)] \label{zero}
\end{equation}
with spacing $2\pi e$ and a more weak logarithmic dependence of real parts. 
Eq. (\ref{zero}) holds to a high accuracy starting from $n=0$.


From the Hadamard factorization, the scattering matrix is constructed as:
\begin{equation}
S(t) = \prod_{n} \frac{t - t_n^*}{t - t_n} = e^{2i\delta},
\end{equation}
ensuring unitarity $|S| = 1$ for real $t$.

In the weak-coupling regime $T \gg T_{K}$, the dominant contribution comes from the nearest Lee--Yang zero pair,
$t_{0},t_{0}^{*} = \sigma_{0} \pm i\tau_{0}$
with $\tau_{0} \simeq 2 \pi e$.
The real part $\sigma_{0}$ may be absorbed into the definition of $T_{K}$, and prefactor modified as the slope of the scaling trajectory to match the high-$T$ limit. Then the relevant distance from the first zero is identified as $t(T) - \sigma_{0}  \simeq L(T)/(2\pi be)$ with $b=\pi [S(S+1)]^{1/2}$, $L(T)= \ln(T/T_{K})$, reflecting  evolution of the coupling constant $g_{eff}(T)$.
Then we have $\delta = \arctan[b/L(T)]$,
\begin{equation}
\frac{\rho_{NS}}{\rho_{u}} = \sin^{2}\delta 
=\frac{b^2}{L^2(T)+b^2 } \label{NS}
\end{equation}
with $\rho_u$ the unitarity limit. This is the Nagaoka-Suhl approximation \cite{NS} which describes Abrikosov-Suhl resonance and yields a correct Born high-$T$ behavior, but  not the  half-height condition $\rho(T=T_K)=\rho_{u}/2$. To fit this condition, interpolation  Hamann's formula \cite{Hamann} is widely used,  
\begin{equation}
\frac{\rho_{H}}{\rho_{u}} 
= \sin^{2}\frac{\delta_{}}{2} 
=\frac 12 \left[1-\frac{L(T)}
{\sqrt{L^2(T)+b^2}}\right], \label{Hamann} 
\end{equation}
which in fact includes $\sqrt{S_{0}}$ rather than $S_{0}$, yielding incorrect high-$T$ asymptotics with a factor of 1/4.

Free energy in the nearest-zero approach is written as 
\begin{eqnarray}
F(T)= - \frac {T}{2b} \left( \arctan\frac {b}{L(T)} 
- \frac {bL(T)}{L^2(T)+b^2} \right)    \label{free} 
    \end{eqnarray}
where the second term in the brackets provides non-automatic cancellation of spurious high-$T$ $1/\ln$-terms in the corresponding Sommerfeld integral over energy $E$, which includes the replacement $\delta(T) \rightarrow \delta(E)$. In other words, $F\sim T[\delta(L)+L\delta'(L)]$, in agreement with the scaling treatment \cite{Wiegmann}. Then we obtain in agreement with \cite{Kondo} for high-$T$ free energy and specific heat
\begin{equation}
F(T)= - \frac{b^2T}{3L^2(T)}, \, C(T) =   \frac{b^2 }{L^4(T)}. \label{free1} 
\end{equation}

At high $T$  the asymptotics  $f(t,0)=tZ(t,0) \simeq 1/\ln|t|-1$ at $t\rightarrow - \infty$ is obtained by the saddle-point method (see Appendix). 
Embedding temperature by a trajectory $t(T)$ such that $\ln(-t)\sim\ln(T/T_K)$, this implies an effective running coupling $$g_{eff}(T)\propto -f(t(T),0)\sim 1/[\ln(T/T_K)+\ln\ln(T/T_K)].$$ Weak-coupling thermodynamics then follows by inserting $g_{eff}(T)$ into the standard perturbative Kondo structures, e.g., $$\delta \chi(T)/\chi_{\rm Curie} \sim g_{eff}(T), \, C(T) \sim S(S+1) g_{eff}^4(T),$$ reproducing the known $1/\ln$ correction to the Curie constant and the $1/\ln^4$ high-$T$ tail of $C(T)$ with numerically important next-order loglog scaling corrections. 

{Thus the logarithmic trajectory of the running coupling constant is built into the analytic structure of  the sophomore’s dream function. We emphasize that the loglog term in $g_{eff}(T)$ is not inserted by hand: it follows from the large-$|t|$ saddle-point asymptotics of $f(t,0)$ derived in Ref. \cite{Irkhin}, and it reproduces, without further adjustment, the known two-loop Kondo scaling function. The equidistant zero ladder of Eq. (\ref{zero}), and hence the entire crossover construction below, is thus a mathematical consequence of matching this one real-axis asymptotic — rather than an independent, \textit{ad hoc} postulate about the complex plane. We note that the saddle-point asymptotics used above to obtain the log+loglog running coupling is derived specifically at $a=0$. For $a>0$ the same asymptotic regime  is instead controlled by a different saddle point and $f(t,a)$ grows exponentially in $|t|$~\cite{Irkhin}; however, $a$ is an irrelevant perturbation in the weak-coupling (high-$T$) limit.}


\textit{Underscreened Kondo model.}
The above results can be also applied to the underscreened Kondo model where $2S$ exceeds the number of scattering channels $n$ (here, we put $n=1$), so that the residual spin $S^* =S-n/2$ retains at $T\rightarrow 0$ \cite{Blandin}. Then the FL regime is not achieved below $T_K$, so that we can use the above approach below $T_K$ too \cite{Wiegmann}.
Thus we write  again the nearest-zero result 
\begin{equation}
\delta(T) = \pi/2 + \arctan \frac{\pi\sqrt{S^{*}(S^{*} + 1)}}{\ln(T/T^{*})}. \label{low}
\end{equation}
where the arctan term comes  from 
the first zero of the residual spin partition function. 
The new energy scale $T^{*} \sim T_K \exp(-A/|j_0|)$ is exponentially small owing to  a small ferromagnetic residual coupling $j_0$, and is expected to be much smaller than $T_K$.
The  resistivity at $T^*\ll T\ll T_K$ follows:
\begin{equation}
\frac{\rho(T)}{\rho_u} = 1 - \frac{\pi^2 S^{*}(S^{*} + 1)}{\ln^2(T/T^{*})} + O(\ln^{-4}).
\end{equation}
The free energy is also obtained from (\ref{free}), (\ref{free1}) by  the replacement $T_K \rightarrow T^*$, $S$ passing smoothly into $S^*$ with lowering $T$, in agreement with \cite{Wiegmann,Coleman}.
At very low $T \sim T^{*}$ we get from (\ref{low}) $F\sim  T /\ln(T/T^*)$, which agrees with the Bethe-ansatz result \cite{Destri}.

\textit{Many-zero approach: Fermi-liquid behavior.}
Let us return to the fully screened case ($S=1/2$). At low and intermediate $T$ we should, generally speaking,  perform summation over all zeros which form in the scaling regime an approximately equidistant ladder. First we use a simple parametrization  $t_{n} \simeq 2\pi e i(n + \alpha)$ neglecting real part of $t_n$. Then, after  scaling $t(T) \rightarrow w = t /(2 \pi e)= T_K/(2\pi T)$, the total phase shift from $N$ zeros reads
\begin{equation}
\delta(w) = \sum_{n=0}^{N-1} \arctan \frac{n + \alpha}{w}
\label{del}
\end{equation}
The offset $\alpha = 1/2$  encodes the universality class of the Kondo fixed point and ensures the Friedel sum rule $\delta(T=0) = \pi/2$.
This sum is structurally identical to a fermionic Matsubara sum. 
Differentiating with respect to $w$ gives
\begin{equation}
\frac{d\delta}{dw} = -\Re\sum_{n=0}^{N-1} \frac{1}{n + \alpha + iw} = \Re[\psi(\alpha + iw) - \psi(N + \alpha + iw)],
\end{equation}
the sum being  represented by the digamma function,
\begin{equation}
\psi(\alpha) = -\gamma + \sum_{n=0}^{\infty}\left( \frac{1}{n+1} - \frac{1}{n+\alpha} \right)
\end{equation}
with $\gamma = 0.577$. Thus $\psi(\alpha + iw)$ may be interpreted as the regularized resolvent of the Lee--Yang zero distribution. Integration then yields
\begin{eqnarray}
\delta(T)& = &\Theta(w) +..., \\ \, \Theta(w)&=&\arg \Gamma(\alpha + iw)=  \Im\ln {\Gamma(\alpha + iw)}.\label{our1}  
\end{eqnarray}
 Note that for $\alpha=1/2$ the function $\Theta(w)$ is related to the  Riemann-Siegel theta function $\vartheta(w)$, 
 \begin{equation}\Theta(w)=2 \vartheta(w) +w\ln(2\pi)+ \pi/4 -\arctan (e^{-\pi w}).
\end{equation}
 Thus Kondo crossover phase shift is essentially the same function that governs zeros of the Riemann zeta function. Also, the relation $\Theta(w)\simeq \Theta(1/w)$ holds in a rather wide region near $w=1$, which suggests an approximate self-duality of the crossover function around $T= T_K/(2\pi)$ ($2\pi T/T_K \rightarrow T_K/2\pi T$).

At low $T$ we meet with a singular behavior, $$\Theta(w) \simeq w \ln w- w-(\pi/2)(1/2-\alpha)$$ at large $w$. To pick up a regular contribution, we can use the Stirling expansion of the Gamma function or apply the Euler-Maclaurin formula  which transforms the discrete sum (\ref{del}) into integral and derivative terms:
\begin{eqnarray} \sum_{n=0}^{N-1} \phi(n) = \int_0^N \phi(x)dx + \frac{\phi(0) + \phi(N)}{2} \nonumber\\+ \sum_{k=1}^p \frac{B_{2k}}{(2k)!} [\phi^{(2k-1)}(N) - \phi^{(2k-1)}(0)] + R_p \end{eqnarray}
 with $B_{2n}$ the  Bernoulli numbers ($B_2 = 1/6$, $B_4 = -1/30$) and $R_p$ a small error term. 
After retaining the  Bernoulli terms and absorbing the remainder into $T_K$ definition we have 
$$\delta = {\pi}/{2} - {1}/{(12w)} +1/(360 w^3) + \cdots,$$
From $\rho/\rho_u = \sin^2\delta$, the low-$T$ expansion yields the  Fermi-liquid correction:
\begin{equation}
\frac{\rho}{\rho_u} = 1 - \frac{\pi^2}{36}\left(\frac{T}{T_K}\right)^2 + O(T^4).
\end{equation}

\textit{Crossover.} 
At intermediate $T$ we should measure the physical phase shift relative to the strong-coupling fixed point,  $\delta_{phys}(T)=\pi/2-\delta(T)$. This is analogous to Wilson’s NRG procedure \cite{Wilson}, where the phase shift is extracted relative to the fixed-point spectrum. 
Choosing the constants to satisfy the half-height condition for the resistivity $\rho = \rho_u \sin^2\delta_{phys}$, the interpolation ultraviolet-regularized formula reads
\begin{equation}
\delta_{phys}(T) = -\Theta(w) +\Theta(w_K) + \pi/4
\label{our}
\end{equation}
where  $w_K = 1/(2\pi)$. At not too low $T$ (and not too small $\alpha$) we can use to a high accuracy the pole approximation $\Gamma(\alpha+iw)\simeq 1/(\alpha+iw)$, $\Theta(w) \simeq -\arg(\alpha+iw) = -\arctan(w/\alpha)$ to obtain after some transformations in a simple algebraic form
\begin{equation}
\frac{\rho }{\rho_u}
= \frac{(1+x_K)^2}{2(1+x_K^2)}\frac{(x+q)^2}{1+x^2}, \,\, x=\frac{T_K}{2\pi  \alpha T}.
\label{ourr2}
\end{equation}
with the asymmetry parameter $ q= {(1-x_K)}/{(1+x_K)}\simeq0.52$ for $\alpha=1/2$. This  Fano‑type shape of the resistivity in the crossover region can be traced to the discrete $\pi/2$ phase increments generated by the infinite Lee–Yang zero ladder. 
Each zero contributes an arctangent resonance; the Euler–Maclaurin smoothing transforms the infinite staircase into a collective pole with residual oscillatory corrections. Subtracting at the matching scale $w_K$  produces interference between the collective background and the discrete correction, yielding the Fano numerator. 
Thus our temperature-space Fano is the Sommerfeld-averaged version of the energy-space Fano resonance observed in STM experiments on Kondo impurities \cite{Kroha,Luo}.


{We emphasize the distinction between our temperature-space Fano shape and (i) the STM energy-space Fano resonance \cite{Kroha,Luo}, which reflects interference between two distinct real-space tunnelling amplitudes, and (ii) the single-pole (Lorentzian) mean-field Kondo resonance ('Kondo cloud') obtained, e.g., in slave-boson mean-field theory, which contains no interference at all. In our construction the asymmetry arises from interference, within the same analytic function $\Gamma(\alpha+iw)$, between the smooth collective contribution of the infinite Lee–Yang zero ladder and the discrete subtraction fixing the boundary condition at the RG matching scale $T_K$ — an interference in coupling-constant (RG) space rather than in real space.}

Eq. (\ref{our}) is readily generalized to the SU($N$) Coqblin-Schrieffer model where $\alpha = 1/N$, and the resistivity is \begin{equation}
{\rho(T)}={\rho(0)}{\sin^2(\delta_{phys})}, \end{equation}
with $\rho(0)=\rho_u\sin^2(\pi/N)$ \cite{Bickers}.

 The crossover for $S>1/2$ underscreened case can be roughly described, in agreement with the Hamann result (\ref{Hamann}) (which can treat to some extent the underscreened situation and depends weakly on $S$ starting from $S=1$), provided that we use in (\ref{our}) the function $\Theta(w,1)= \Im\ln {\Gamma(1 + iw)}$.

The regularization  in (\ref{our}) plays the role of Wilson's discretization which encodes the crossover physics.  
Note that a simple  Wilson-type regularization $$\delta_{phys}(T) = -\Im\ln[{(\Gamma(\alpha + iw)}/(\Gamma(\alpha + iw)-\Lambda\Gamma(1+\alpha + iw))]$$
yields, with account of the half-height condition, 
$\rho/ \rho(0)= 1/(1+(T/T_K)^2)$, which is a poor result, except for the large-$N$ case.

\textit{Drift corrections.}
The real parts of the Lee-Yang zeros should result in a slight modification of our picture. The $n$-logarithmic dependences of $\Re t_n$ produce logarithmic drift corrections to the phase sift. Writing down the sum
$$\Delta\delta(w)=\sum_n[\arctan(\xi_n/(w+\Delta_n)-\arctan \xi_n/w]$$ with $\Delta_n=-(1/4\pi)\ln(2\pi^2\xi_n)$, $\xi_n=n-1/8$, expanding in $\Delta_n$, passing from summation to integration over $\xi$  and regularizing the integral we obtain the correction to the phase shift  as 
\begin{equation}
\Delta\delta_{phys}(T)= -\frac 18 \ln ( T_K/T), \label{drift}
\end{equation}
This correction is irrelevant  in the FL region, but influences the behavior in the crossover. 

Thus in our approach the screening is built up by accumulating scattering resonances (Lee-Yang zeros), exact summation of  infinitely many non-interacting roots being provided by the Gamma function. This picture is somewhat similar to that in Bethe ansatz and thermodynamic Bethe ansatz (TBA) where the resonances (Bethe roots) interact with each other through a kernel $K=1/(2\cosh x )$ \cite{Zamolodchikov,Tongeren}.  Our uniformly-spaced zeros corresponds to Bethe roots in the non-interacting limit. This interaction slightly reshuffles the resonance positions but doesn't change the total phase shift. Our logarithmic correction  captures the dominant effect of zero-zero interactions in the coupling-constant plane, just as the TBA kernel captures root-root interactions in rapidity space.




\begin{figure}
	\includegraphics[width=0.49\textwidth]{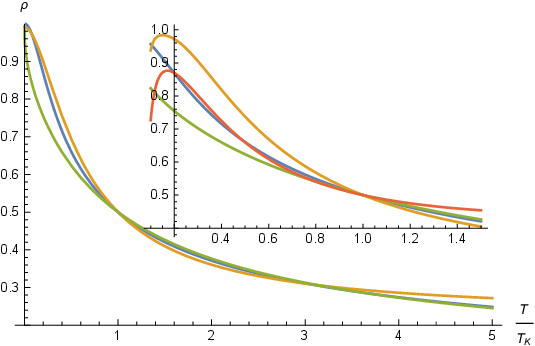}
		\caption{Temperature dependence of the relative resistivity $\rho(T)/\rho_u$ according to (\ref{our}) (orange) in comparison with the Hamann approximation (\ref{Hamann}) (green) and the NRG-interpolating approximation (\ref{gold}) (blue). The inset shows the low-$T$ behavior;  the result with drift correction improving the  agreement with (\ref{gold}) is shown by red. The cutoff $1/w\rightarrow 1/w+\exp(-T/T_K)$ is used for the main figure, but not for inset.
	}
	\label{22}       
\end{figure}

A comparison with previous results for resistivity is presented in Fig.~\ref{22}.
We show the results of the  Hamann approximation (\ref{Hamann}) and   empirical fitting  \begin{equation}
\rho(T)/\rho_u =(1 + (2^{1/s}-1)(T/T_K)^2 )^{-s} \label{gold}
\end{equation}
with $s=0.22$ \cite{Goldhaber}, which provides a very good agreement with NRG \cite{Costi} at not too high $T$, including the low-$T$ region (where the Hamann approximation is wrong).
In the infrared region (below $T_K/2 \pi$), the formula (\ref{our}) still requires regularization to suppress low-$T$ oscillations,  e.g., $w= 1/(2\pi T+ \exp (-T/T_K))$.
At high $T>5 T_K$, Eqs.~(\ref{our}), (\ref{ourr2}), although do not yield the correct asymptotic Born behavior (\ref{NS}), seem to work better than two other approaches which yield too rapid decrease with increasing $T$. Thus Eq.(\ref{ourr2}) provides a promising alternative for the Hamann's approximation which is widely used when treating experimental data. 


\textit{Thermodynamic properties.} 
Description of thermodynamic properties in our approach turns out to be more difficult.
Applying the Hadamard product (\ref{Had})  to the free energy $F=-T \ln Z(t,a)$ we derive similar to (\ref{our})
\begin{equation}
F(T) = - T \Re \ln {\Gamma(\alpha + iw)} +...  \label{free2}
\end{equation}
Using the reflection formula $\Gamma(z)\Gamma(1-z)=\pi/\sin \pi z$, we have  $F(T)=  T\ln (\cosh\pi w)/2$ for $\alpha =1/2$ and $F(T)=  T\ln (\sinh\pi w)/2$ for $\alpha =1/N \rightarrow 0$, which is a too simple and rather poor structure. Indeed, we derive for $\alpha =1/2$
a Schottky-like result for specific heat,  $$C(T)= T_K^2/({8T^2\cosh^2(T_K/2T)}),$$
which provides a qualitatively correct Kondo behavior with a maximum at $T \simeq 0.4T_K$, but does not reproduce low-$T$ FL behavior. 

{This limitation can be understood as a consequence of the fact that our formalism controls only the imaginary (on-shell, phase-shift) channel of the effective T-matrix, $\Im \ln \Gamma (\alpha+iw)$, while thermodynamics additionally requires the properly Kramers–Kronig-related real part. A structurally similar Im/Re asymmetry is well known from the non-crossing approximation for the Anderson/Kondo $T$-matrix, which reproduces resistivity well but violates Fermi-liquid Ward identities unless vertex (conserving) corrections are included \cite{Bickers}. A consistent thermodynamic extension of the present approach likely requires promoting the phase-shift (digamma) level of description to the dilogarithm level familiar from the thermodynamic Bethe ansatz.}

 To improve the situation, we can  calculate the free energy in terms of our phase shift -- in spirit of the  Nozieres approach \cite{Nozieres} from the Sommerfeld integral 
\begin{eqnarray}
F(T,H)=-\frac{2T}{\pi}\int_0^\infty d\epsilon \delta\left(\sqrt{T^2+H^2+{\epsilon^2}/{\pi^2}}\right) \nonumber \\
\times\ln\left(1+e^{-\epsilon/T}\right),  \label{free3}
    \end{eqnarray}
Magnetic susceptibility calculated according to (\ref{free3})  provides qualitative agreement with known results (a maximum at about  $0.5 T_K$).

\textit{Effects of spin dynamics: Kondo lattices.} 
When treating concentrated Kondo systems, a competition with intersite  exchange  interactions becomes important.  This can be taken into account by the replacement $T \rightarrow \sqrt{T^2+\omega_{sf}^2}$ in the Kondo singular terms (e.g., in Eqs.(\ref{free}), (\ref{free2})) with $\omega_{sf}$ the spin-fluctuation frequency.  
A more consistent treatment can be performed in the scaling approach \cite{IK,UFN} which provides a pair of scaling equations for the mutually renormalized coupling $g_{eff}(T)$ and $\omega_{sf}(T)$. 
Thereby, the running effective Kondo scale is defined  using the running coupling:
\begin{equation}
T_K^{eff}(T)=T\exp(-1/g_{eff}(T))
\end{equation}
which reduces to the standard one-impurity results in the absence of spin dynamics, but can be considerably renormalized during a scaling (temperature) trajectory. Then our above results can be modified by introducing the quantity
\begin{equation}
w(T)=\frac{T_K^{eff}(\max(T,T_0))}{2\pi\sqrt{T^2+\omega_{sf}^{2}(\max(T,T_0))}}
\end{equation}
with $T_0$ the temperature where scaling stops.

\textit{Anderson model.}
For the Anderson impurity model, we can use the substitution $w(T)=T_K/(2\pi T_{eff})$ with
$T_{eff}=\sqrt{T^2+E_c^2}, \quad E_c\simeq \min(|\epsilon_f|,\epsilon_f+U,\Gamma)$
where $\epsilon_f$ and $\Gamma=\pi N(E_F)V^2$ are the position (relative to $E_F$) and width of the $f$-level hybridized with conduction band. In the asymmetric (mixed valence) case we introduce $a\sim \pi(\epsilon_f+U)/\Gamma$. To satisfy the Friedel sum rule $\delta(0) = \pi n_f/2$ for $\alpha=1/2$, we have to add the potential scattering term $\delta_p= \pi (n_f -1)/2$, e.g., determined by the Hartree approximation, 
into (\ref{our}) and into the nearest-zero result which is similar to (\ref{NS}): 
\begin{equation}
\rho(T)/\rho(0)= \frac{(b\cos\delta_p + L(T_\text{eff})\sin\delta_p)^2} {L^2(T_\text{eff}) + b^2}.
\end{equation}

\textit{Conclusions.}
We have presented an analytical framework for the Kondo crossover
based on the analytic structure of the renormalization-group flow. 
A main insight is that  scaling implies an approximately equidistant ladder of Lee–Yang zeros in the complex coupling-constant plane. 
Our partition function  $Z$ provides an effective   model intended to approximate the impurity contribution to the path integral after the conduction electrons have been integrated out. 
The method offers not a replacement for Bethe ansatz or NRG, but a complementary analytic lens that exposes the deep connection between renormalization-group flow, complex-zero distributions, and the Gamma function.
In this interpretation the integration variable  parametrizes progress along the screening trajectory (associated with the buildup of the scattering phase shift), while the microscopic RG flow enters through a temperature-dependent embedding. 

The central assumption is that the analyticity and zero structure, in particular, its Lee–Yang zero ladder and Hadamard factorization capture the correct global analytic organization of the crossover, while the detailed microscopic model enters primarily through the mapping  and the parameters controlling the zero ladder. 
Differences in accuracy between observables provide a diagnostic for systematic improvements, such as 
incorporating corrections to the ideal ladder (e.g., logarithmic drifts or effective zero–zero interactions) as fluctuations around the reduced description.

\textit{Acknowledgment}. The investigation was performed with support by Ministry of Science and Higher Education of the Russian Federation.

{\textit{Appendix. Properties of the function $f(t,a)=tZ(t,a)$}.}
Here we review some results of Ref. \cite{Irkhin}. We have
{\begin{equation}
	f(t,a)=t\int_{0}^{\infty }\exp(te^{-y}(y-\ln \tilde{a}-y)dy. \label{y}
\end{equation}%
where $\tilde{a} = e^a$ (used in Ref. \cite{Irkhin}). For large $t>0$, this exponential  integral can be estimated by the Laplace method, i.e., by expanding the exponent near the  point where its derivative vanishes, $y_0\simeq 1+\ln \tilde{a}$. Putting $y=y_0+\delta $ and expanding in $\delta $ we get
\begin{equation*}
	te^{-y}(y - \ln \tilde{a} )\simeq (t/e\tilde{a})(1-\delta ^{2}/2).
\end{equation*}%
Evaluating the integral over $\delta$ in the infinite limits we obtain the exponential asymptotics%
\begin{equation}
	f(t,a)=\left( \frac{2\pi t}{e \tilde{a}}\right) ^{1/2}\exp \frac{t}{e \tilde{a}}.
	\label{f16}
\end{equation}%

For $\tilde{a}<1$ ($a<0$),  large $|t|$ and $t<0$ the integral (\ref{y}) is determined by the vicinity of the saddle point which is defined by the equation
\begin{equation}
	\frac{d}{dy}(te^{-y}(y-\ln \tilde{a})-y)=te^{-y}(1+\ln \tilde{a} -y)-1=0.
		\label{eqs}
\end{equation}
Solving  this equation by iterations, we can approximate the integral by putting $y \simeq  \ln(-t)+\ln (\ln (-t/\tilde{a})-1)...$ in the brackets of the exponent,
\begin{equation}
	f(t,a)\simeq t\int_{0}^{\infty} \exp ( te^{-y} [\ln(-t/\tilde{a})+\ln(\ln (-t/\tilde{a})-1) ]-y )dy. \label{ys}
\end{equation}%
Changing  the variables $x=e^{-y}$ we can perform integration to obtain
\begin{equation}
	f(t,a)\simeq -1/[\ln(-t/\tilde{a})+\ln (\ln (-t/\tilde{a})-1)].
	\label{fas}
\end{equation}
For the marginal situation $a=0$ ($\tilde{a}=1$) we have 
\begin{equation}
	f(t,0)\simeq -1/[\ln(-t)+\ln (\ln (-t)-1)]-1.
	\label{fas1}
    \end{equation}

An interpolation expression for  $f$, working in the complex plane, 
can be written in terms of the error function:
\begin{eqnarray}
	f(t,a) =\frac {\varphi(t\tilde{a}/e)} {\tilde{a}}	\exp \left(\frac{t}{e \tilde{a}}\right), \label{77}
	\\
	\varphi(x)=\left( \frac{2\pi x}{e}\right) ^{1/2} {\rm erf}(x ^{1/2})
	\label{ha11}.
\end{eqnarray}%
Using the results of Ref. \cite{error} we obtain the expression for the poles (\ref{zero}).


\begin{thebibliography}{99}

\bibitem{Kondo}
J. Kondo, \textit{Theory of Dilute Magnetic Alloys}, Solid State Physics, Vol. 23, edited
by F. Seitz, D. Turnbull and H. Ehrenreich (New York: Academic Press, 1969), p. 183.  

\bibitem{Hewson} A.C. Hewson, \textit{The Kondo Problem to Heavy Fermions} (Cambridge, 1997).


\bibitem{Andrei}
N. Andrei, K. Furuya, J.H. Lowenstein, Solution of the Kondo problem, Rev. Mod. Phys. \textbf{55}, 331 (1983).

\bibitem{Wiegmann}
A.M. Tsvelick, P.B. Wiegmann, Exact results in the theory of magnetic alloys, 
Adv. Phys. \textbf{32}, 453 (1983).

\bibitem{Wilson}
K.G. Wilson, Critical phenomena and the Kondo problem, Rev. Mod. Phys. \textbf{47}, 773 (1975).

\bibitem{Bulla}
R. Bulla, T. Costi, T. Pruschke, Numerical renormalization group method for quantum impurity systems,  Rev. Mod. Phys. \textbf{80}, 395 (2008).

\bibitem{Hamann}
D. R. Hamann, New solution for exchange scattering in dilute alloys, Phys. Rev. \textbf{158}, 570 (1967).

\bibitem{Nozieres}
P. Nozi\`eres, A ``Fermi-Liquid'' description of the Kondo
problem at low temperatures, J. Low Temp. Phys. \textbf{17}, 31 (1974).

\bibitem{Gabor} Gabor Roman, Extension of the Sophomore’s Dream, An. st.  Univ. ``Ovidius'' Constanta. Seria Matematica, \textbf{29},  211 (2021).


\bibitem{Irkhin} V. Yu. Irkhin, Sophomore’s dream function: asymptotics, complex plane behavior and relation to the error function, arXiv:2501.10936 [math.CA].


\bibitem{Bernoulli}
Johann Bernoulli, Demonstratio methodi analyticæ, qua determinataest aliqua quadratura exponentialis per Seriem, Opera omnia, vol. 3 (1697).

\bibitem{Lee}
C. N. Yang and T. D. Lee, Statistical Theory of Equations of State and Phase Transitions. I. Theory of Condensation, Phys. Rev. \textbf{87}, 404 (1952).

\bibitem{Fisher}
M.E. Fisher, \textit{The nature of critical points}, in Lectures in Theoretical Physics 7C, 1 (1965).

\bibitem{Boo}
X. Peng, H. Zhou, Bo-Bo Wei, J. Cui, J. Du, Ren-Bao Liu, Observation of Lee-Yang zeros, Phys. Rev. Lett. \textbf{114}, 010601 (2015).

\bibitem{Stephanov}
M.A. Stephanov, QCD critical point and complex chemical potential singularities, Phys. Rev. D\textbf{73}, 094508 (2006).

\bibitem{Ding}
H.-T. Ding, S. Mukherjee, P. Petreczky, K.-F. Ye, QCD Chiral Crossover line from Lee-Yang edge singularities, arXiv:2608.05752.


\bibitem{Yuval}
P. W. Anderson, G. Yuval, and D. R. Hamann,
Exact results in the Kondo problem. II. Scaling theory, qualitatively correct solution, and some new results on one-dimensional classical statistical models, 
Phys. Rev. B\textbf{1}, 4464 (1970).

\bibitem{NS} Y. Nagaoka, Self-consistent treatment of Kondo's effect in dilute alloys. Phys. Rev. \textbf{138}, A1112 (1965).

\bibitem{Blandin}
 P. Nozieres and A. Blandin,   Kondo effect in real metals. J. de Phys. \textbf{41}, 193 (1980).

\bibitem{Coleman}
P. Coleman and C. Pepin, Singular Fermi liquid behavior in the underscreened Kondo model, Phys. Rev. B\textbf{68} 220405 (2003).

\bibitem{Destri}
N. Andrei and C. Destri, Solution of the multichannel Kondo problem, Phys. Rev. Lett. \textbf{52}, 364 (1984).

\bibitem{Kroha}
O. Ujsaghy, J. Kroha, L. Szunyogh, A. Zawadowski,
Theory of the Fano resonance in the STM tunneling density of states due to a single Kondo impurity,
Phys. Rev. Lett.\textbf{ 85}, 2557 (2000).

\bibitem{Luo}
H. G. Luo, T. Xiang, X. Q. Wang, Z. B. Su, L. Yu,
The Fano resonance for Anderson impurity systems, Phys. Rev. Lett. \textbf{92}, 256602 (2004).


\bibitem{Bickers}
N. E. Bickers, Review of techniques in the large-N expansion for dilute magnetic alloys, Rev. Mod. Phys. \textbf{59}, 845 (1987).

\bibitem{Zamolodchikov} 
Al.B. Zamolodchikov, Thermodynamic Bethe ansatz in relativistic models, Nucl. Phys. B 342, 695 (1990).

\bibitem{Tongeren}
S. J. van Tongeren, Introduction to the thermodynamic Bethe ansatz, J. Phys. A \textbf{49}, 323005 (2016). 

\bibitem{Goldhaber}
D. Goldhaber-Gordon, J. Goeres, M. A. Kastner, Hadas Shtrikman, D. Mahalu, U. Meirav, From the Kondo Regime to the Mixed-Valence Regime in a Single-Electron Transistor, 
Phys. Rev. Lett. \textbf{81}, 5225 (1998).

\bibitem{Costi}
T.A. Costi, A.C. Hewson, V. Zlatic, Transport coefficients of the Anderson model via the numerical renormalization group, J. Phys.: Condens. Matter \textbf{6}, 2519 (1994).



\bibitem{IK}
V.Yu. Irkhin, M.I. Katsnelson, Scaling picture of magnetism formation in the anomalous f-electron systems:
Interplay of the Kondo effect and spin dynamics, Phys. Rev. B \textbf{56}, 8109 (1997).

\bibitem{UFN}
V.Yu. Irkhin, Unusual magnetism of the Kondo lattice, Phys. Usp. \textbf{60}, 747 (2017).

\bibitem{error}
H. E. Fettis, J, C. Caslin and K, R. Cramer,
Complex zeros of the error function and of
the complementary error function,
Mathematics of Computation \textbf{27}, 401 (1973).


\end{thebibliography}
\end{document}